\documentclass[fleqn,10pt]{wlscirep}
\usepackage{amsmath}
\usepackage{amsfonts}
\usepackage{amssymb}
\usepackage{color}
\usepackage{graphicx}
\usepackage{array}
\usepackage{graphicx,epsfig}
\usepackage{epstopdf}
\DeclareGraphicsExtensions{. jpg,. pdf, . mps, .png, .eps, . ps, . EPS}

\def\bc{\begin{center}}
\def\ec{\end{center}}
\def\bea{\begin{eqnarray}}
\def\eea{\end{eqnarray}}

\newcommand{\Avg}[1]{\left\langle{#1}\right\rangle}

\begin{document}

\title{Complex Network Geometry and Frustrated Synchronization}

\author[1]{Ana P. Mill\'an} 
\author[1]{Joaqu\'in J. Torres}
\author[2]{Ginestra Bianconi}  
\affil[1]{Departamento de Electromagnetismo y F\'isica  de la Materia
and Instituto Carlos I de F\'isica Te\'orica y Computacional, Universidad de Granada, 18071 Granada, Spain}
\affil[2]{School of Mathematical Sciences, Queen Mary University of London, E1 4NS London, United Kingdom}

\begin{abstract}The dynamics of networks of neuronal cultures  has been recently shown  to be strongly dependent on the network geometry and in particular on their  dimensionality.  However, this phenomenon has been so far mostly unexplored from the theoretical point of view. Here we reveal the rich  interplay between network geometry and  synchronization of coupled oscillators in the context of a simplicial complex model of manifolds called Complex Network Manifold. The networks generated by this model  combine  small world properties (infinite Hausdorff dimension) and a  high modular structure with   finite and tunable spectral dimension. We show that the networks display  frustrated synchronization for a wide range of the coupling strength of the oscillators, and that the synchronization properties are directly affected by the spectral  dimension of the network.  
 \end{abstract} \maketitle

\section{Introduction}

Network Theory has provided a large evidence that network topology affects the dynamics and function of complex networks \cite{BA,SW,Doro_book,Boccaletti_review,Newman_book,Laszlo_book,Santo}. Recently, growing attention has been devoted to analyze networks not only from the topological viewpoint but, also, from a geometrical perspective \cite{perspective,Marc,Evans,SB3}.
Characterizing the geometry of complex networks is particularly relevant for brain research \cite{Bullmore,Bassett}, where the embedding in a three-dimensional space is essential to understand the wiring diagram characterizing the connections between brain regions --or connectome \cite{Zoltan,Petra,Sporns}-- and also at a more microscopic neuronal level. 
Interestingly, recent experimental results have shown that the synchronization properties of neuronal cultures grown on 2D slices differ considerably from those grown on 3D scaffolds; these last ones turn out to be much more likely to maintain synchrony and have two regimes: a highly synchronized regime and a moderately synchronized regime \cite{Torre}. This result can be compared with results obtained in the framework of the Blue Brain project where is has been observed that pairs of neurons have a more significant correlations in their dynamics if they belong to higher dimensional simplicies \cite{BB}.
 
These results reveal that not only network topology but also network geometry plays a crucial role in determining the dynamical properties of a network.
However, the connection between network geometry and dynamics has been so far mostly unexplored. Here, we provide a theoretical framework based on simplicial complexes \cite{Bassett,Emergent,Hyperbolic,NGF,CQNM,Complexity,Vaccarino2} to investigate  the effect of network geometry on synchronization dynamics. 
Our study reveals that a finite spectral dimension of geometrical networks can be combined with a complex modular network structure to give rise to frustrated synchronization and spatio-temporal fluctuations of the order parameter of the synchronization dynamics.
In this way, we provide a clear evidence of the important effect that network geometry has on the dynamics of complex networks.

Synchronization phenomena are the subject of extensive research in physical, biological, chemical and social systems \cite{Kurths,Arenas}. The synchronization properties of a network are influenced by the network topology \cite{Boccaletti} and by its spectral properties \cite{Barahona,Donetti}. For instance the stability of a fully synchronized state is known to depend crucially on the ratio between the Fiedler eigenvalue  (i.e. the smallest non zero eigenvalue) and the largest eigenvalue of Laplacian \cite{Barahona,Donetti}. 

Interestingly, complex network topologies characterized by a hierarchical modular structure have been shown to display a dynamical phase called {\em frustrated synchronization} \cite{Moretti} in which the order parameter at large times is not stationary but it is instead  affected by large spatio-temporal fluctuations. This exotic phase occurs in highly modular networks where contact processes have also a dynamical behavior dominated by rare regions of activity in correspondence with the so-called Griffith phase \cite{Villegas,safari,Cota}. However, in Ref. \cite{Moretti} the considered network structures have a finite Hausdorff dimension while a large number of brain networks is known to be small-world, \cite{SW,Bullmore} (i.e. have an infinite Hausdorff dimension). Therefore exploring whether  such dynamical phase can appear in small world networks, such as those that characterize the structure of the brain, has particular relevance in the field of neuroscience and neural computation.

Here, we show that it is possible to generate network geometries called Complex Network Manifolds \cite{CQNM,NGF,Hyperbolic,Complexity} combining both modular hierarchical structure and a small-world property of the network (i.e. Hausdorff dimension $d_H=\infty$). 
Interestingly, Complex Network Manifolds tessellate $D$ dimensional spaces and the dimension $D$, together with their modular structure, strongly affect their synchronization properties. As a result of their complex network geometry, these network structures can sustain frustrated synchronization for a large range of their parameter values, with  the stability of this phase depending crucially on the dimension $D$. 
The observed spatio-temporal fluctuations of the order parameter appear when the dynamics is strongly affected by localized eigenvectors, identifying "rare" regions of activity, and driving global oscillations of the synchronization global order parameter.

\section{Complex Network Manifolds}

\subsection{Definition and basic structural properties}

Simplicial complexes are natural objects to be considered when investigating network geometry. In fact, intuitively they can be interpreted by geometrical network structures built by geometrical "LEGO blocks" (the simplices). 

A $d$-dimensional simplicial complex is formed by $d$-dimensional simplices (fully connected networks of $d+1$ nodes) such as nodes ($d=0$), links $(d=1)$, triangles $(d=2)$ , tetrahedra $(d=3)$ etc., glued along their faces.
Here by a face of a $d$-dimensional simplex, we indicate a $\delta$-dimensional simplex with $\delta<d$ formed by a subset of its nodes.
A simplicial complex has the additional property that if a simplex $\alpha$ belongs to the simplicial complex $\mathcal K$  (i.e. $\alpha\in {\mathcal K}$) then also all its faces $\alpha'\subset \alpha$ belong to the simplicial complex $\mathcal K$ (i.e. $\alpha'\in {\mathcal K}$).

Complex Network Manifolds (CNM) \cite{CQNM,NGF,Hyperbolic} are generated by a non-equilibrium growing network dynamics.
Specifically, Complex Network Manifolds are  growing simplicial complexes of dimension $d$. 
CNM are discrete manifolds  generated by gluing subsequently $d$-dimensional simplices along their $(d-1)$ faces.
Every $(d-1)$ face $\alpha$ of the CNM is characterized by an incidence number $n_{\alpha}$ indicating the number of $d$-dimensional simplices incident to it minus one.
Initially (at time $t=1$), the CNM is formed by a single $d$-dimensional simplex.
At any subsequent step (at time $t>1$) a new $d$-dimensional simplex is glued to a $(d-1)$- face $\alpha$ with probability
\bea \Pi_{\alpha} = \frac{ 1-n_\alpha}{\sum_{\alpha'}(1-n_{\alpha'})}.\eea
In Ref. \cite{Hyperbolic} the exact degree distribution of CNMss has been analytically  derived.  Mainly, the degree distribution is an exponential degree distribution for dimension $d=2$ and a power-law for dimension $d>2$ with power-law exponent $\gamma$ given by 
\bea 
 \gamma=2+\frac{1}{d-2}.
 \eea
 The resulting network is small-world and has an infinite Hausdorff dimension (see Supplementary Information).

\begin{figure}[!h] \center \includegraphics[scale=0.4]{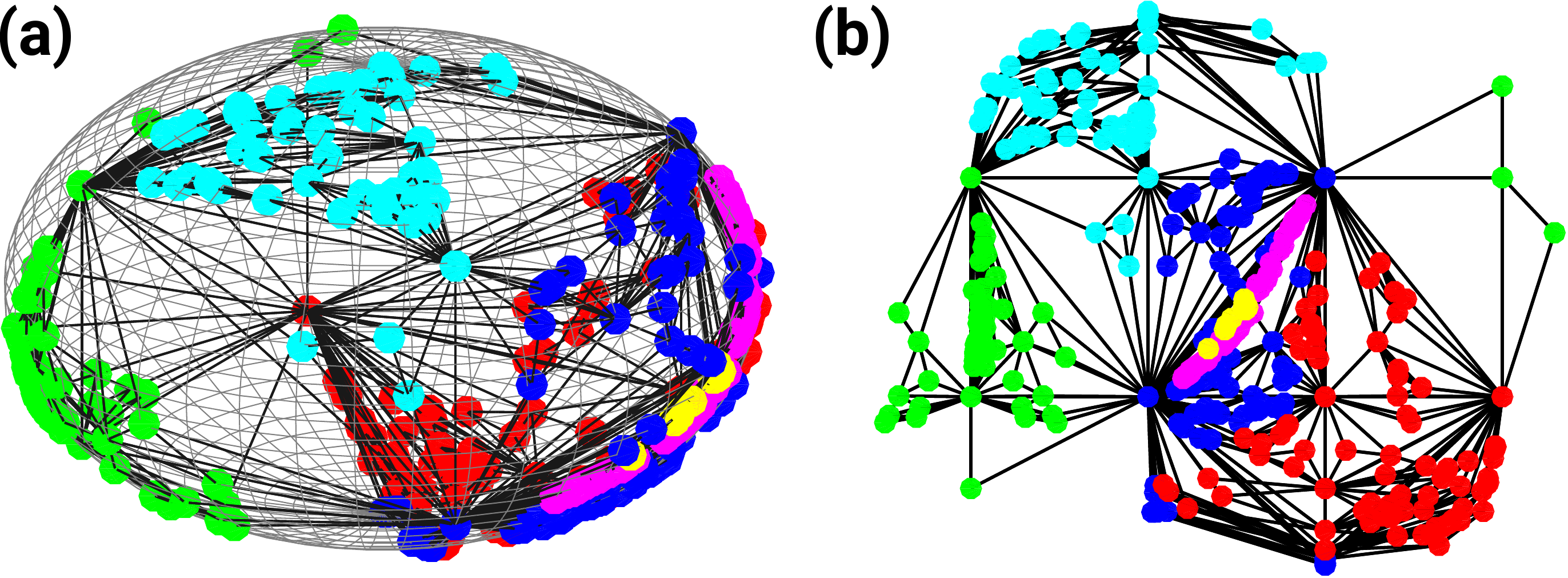} \protect\caption{{\bf Complex Network Manifold's dimensions.} The Complex Network Manifold constructed from $d$-dimensional simplicies with $d=3$ can be interpreted either (panel (a)) as a $d=3$ dimensional manifold with boundary (topologically equivalent to a sphere) or alternatively as a $D=2$ dimensional manifold without boundary tessellating the $D=2$ dimensional surface of the sphere. Panel (b) represents the projection of the $D=2$ manifold on the plane with the Cartesian coordinates indicating the azimuth and elevation of point on the surface of the sphere,respectively. Note that for better visualization purposes we have omitted the links connecting nodes at the opposite sides of the plane. The node colors indicate different communities detected using the Gen-Louvain algorithm \cite{Gen_louvain,Louvain}.\label{fig:NEW_graph_plot} } \end{figure}

\subsection{Network dimensions}
Complex Network Manifolds can be considered as manifolds of dimension $d$ with all the nodes placed on the boundary of the manifold. Alternatively, CNMs formed by $d$-dimensional simplices can be interpreted as $D=d-1$ dimensional manifolds without boundary.

For instance, if we consider a CNM of dimension $d=3$, we can embed the manifold in a $d=3$ dimensional sphere of radius one without any crossing of the faces. To this end, the first tetrahedron is a regular tetrahedron inscribed into the $3$-dimension sphere of radius one, and every new tetrahedron has the new node on the surface of the sphere and shares a triangular face with the already existing simplicial complex. Since in this embedding all the nodes of the CNM are on the $D=2$ dimensional surface of the sphere, the network can be also interpreted as a tessellation of the $D=2$ dimensional manifold formed by the surface of the $d=3$ sphere (see Figure $\ref{fig:NEW_graph_plot}$).

In this $D=2$ triangulated space, the elementary move of adding a new tetrahedron corresponds to the placement of a new node in a middle of a randomly selected triangle of the $D=2$ triangulation and the establishment of three new links between the new node and each of the nodes of the selected triangle.  Most notably, this construction reveals that CNMs in $d=3$ are random Apollonian Networks \cite{apollonian1,apollonian2,apollonian3}.

Similarly, a CNM of dimension $d=4$ can be interpreted either as a $4$-dimensional manifold with boundary or as the tessellation of a $D=3$ space. In this case, the addition of a new $4$-simplex correspond to the selection of a tetrahedron forming the tessellation of the $D=3$, the placement of a new node in the middle of it and the establishment of four new links between the new nodes and each of the nodes of the selected tetrahedron. 

Therefore, we can naturally associate both, dimension $d$ and dimension $D$ of its natural embedding spaces to the CNM formed by simplices of dimension $d$.
 
A further dimension of any CNM is its spectral dimension characterizing the property of its Laplacian spectrum. This dimension will be characterized in the next paragraph.

\subsection{Spectral and localization properties}
The spectral properties of complex networks have been shown to be particularly relevant to reveal the interplay between network structure and synchronization dynamics on the network.  Here, we emphasize the relationship between spectral properties and network geometry. To this end, we characterize the spectral properties of the normalized Laplacian matrix ${\bf L}$ characterizing any given CNM. The normalized Laplacian ${\bf L}$ has elements
\bea L_{ij}=\delta_{ij}-a_{ij}/k_i,\eea
where $a_{ij}$ define the adjacency matrix (with $0$ entries for non connected nodes and $1$ entries for connected ones) and $k_i$ degree of node $i$ in the associated network skeleton.  The normalized Laplacian has real eigenvalues $0=\lambda_1\leq \lambda_2 \leq \ldots \leq \lambda_N$.

\begin{figure}[!h]
\center
\includegraphics[scale=0.5]{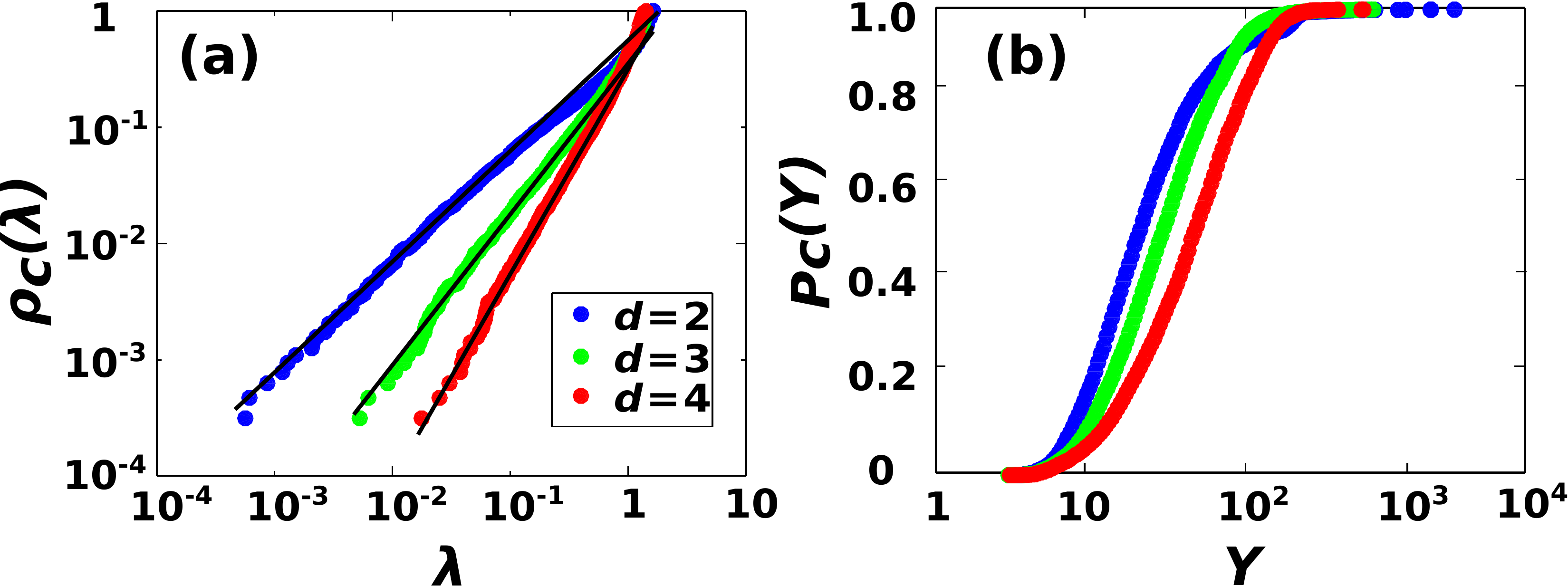}
\protect\caption{{\bf Spectral properties of Complex Network Manifolds.} Panel (a) shows the cumulative  distribution of eigenvalues $\rho_c(\lambda)$ for CNMs of dimension $d=2,3,4$.  Solid (black) lines indicate the power-law fit of
$\rho_c(\lambda)$. Panel (b) shows the cumulative distribution $P_c(Y)$ of the participation ratio $Y$ of the CNM modes (legend is the same as in panel (a)). Data are extracted from a single realization of CNMs of size $N=6400$.\label{fig:spectra_ALL} }
\end{figure}

A large number of complex networks are characterized by a \emph{spectral gap}, i.e. their second smallest (so-called Fiedler) eigenvalue $\lambda_2$ does not approach zero as the network size grows. On the other hand, CNMs, like regular lattices, have a spectral gap that approaches zero for large network sizes (see figure $\ref{fig:spectra_ALL}$). This ensures that one can define the {\it spectral dimension} $d_S$, which characterizes the power-law scaling of the density of eigenvalues $\rho(\lambda)$ for $\lambda\ll 1$ as \cite{Toulouse,Burioni-Cassi1996},
\bea \rho(\lambda)\simeq \lambda^{d_S/2-1}.\eea 
$d_s$ can be calculated starting from the associated cumulative distribution $\rho_c(\lambda)$ determining the probability that a random eigenvalue is less than $\lambda$, i.e.
\bea\rho_c(\lambda)=\int_0^{\lambda} d\lambda' \rho(\lambda')~\simeq \lambda^{d_S/2}.\eea

For finite $D$-dimensional lattices, it holds that $d_S=D$ (where $D$ is both the Euclidean and the Haussdorf dimension) but, in general, the spectral dimension does not have to coincide with the Hausdorff dimension of the network nor with the dimension of the space it tessellates \cite{Emergent,Toulouse}.

Remarkably, CNMs formed by $d$-dimensional simplices have spectral dimension \bea d_S\simeq d \eea as we have checked numerically for dimensions $d=2, 3, 4$. Indeed, in Figure $\ref{fig:spectra_ALL}$(a), we show the cumulative eigenvalue distribution $\rho_c(\lambda)$ and the fitted power-law behavior for small $\lambda$, which yields $d_S=2.00(2)$ for $d=2$, $d_S=3.02(4)$ for $d=3$, $d_S=3.96(8)$ for $d=4$.

The eigenvectors of the normalized Laplacian are also useful to reveal relevant properties of the CNMs. Given that the normalized Laplacian is an asymmetric matrix, it is characterized by a set of left ${\bf u}^{\lambda}_{L}$ and a set of right eigenvectors ${\bf u}^{\lambda}_{R}$ which, in general, do not coincide. Left and right eigenvectors associated with the same eigenvalue are normalized according to the condition 
\bea \sum_{i=1}^N u_{i,L}^{\lambda}u_{i,R}^{\lambda'}=\delta_{\lambda,\lambda'}.  \eea
The localization of any given eigenvector can be quantified using the {\em participation ratio} $Y$, which is an indicator of the number of nodes on which such an eigenvector has a significantly different from zero value, i.e.  
\bea Y=\left[\sum_{i=1}^N \left(u_{i,L}^{\lambda}u_{i,R}^{\lambda}\right)^2\right]^{-1}.  \eea
In Figure $\ref{fig:spectra_ALL}$(b), we show that a very large fraction of eigenvectors are localized on a small fraction of nodes, as indicated by a small value of their participation ratio $Y$ compared with the total number of nodes of the network, $N$. 
This differs from what happens in regular lattices, where eigenfunctions (oscillation modes) are typically delocalized.

\section{Results}

\subsection{Synchronization }
Here we investigate the synchronization properties of coupled oscillators --defined as in the Kuramoto model \cite{kuramoto1975self,strogatz2000kuramoto,acebron2005kuramoto,Kurths,Arenas} -- running on top of CNMs as a function of the network dimensionality.
To each network node $i$, we associate a non-linear oscillator whose phase $\theta_i$ changes in time according to its own internal frequency $\omega_i$ and the coupling to other connected oscillators, obeying
\bea
\dot{\theta}_i(t)=\omega_i + \sigma\sum_{j=1}^N  \frac{a_{ij}}{k_i} \sin (\theta_j - \theta_i),
\label{Kuramoto}
\eea
where $\sigma$ is the control parameter tuning the overall coupling strength, and the internal frequencies $\omega_i$ are random variables independently drawn from a normal distribution with mean $0$ and variance $1$, i.e. ${\mathcal N}(0,1)$\cite{Note}.  In order to attenuate the effect of possible heterogeneities in the degree distribution --and to emphasize that of network geometry-- in Eq. $(\ref{Kuramoto})$ the effective coupling between node $i$ and its neighbours has been normalized by its degree $k_i$.  In order to quantify the degree of synchronization, we employ the standard Kuramoto order parameter, defined as \bea Z(t)=R(t)e^{i\phi (t)} = \frac{1}{N}\sum_j e^{i\theta_j (t)}, \eea where $R(t) \in[0,1]$ is a real variable (function of $t$) that quantifies the level of global synchronization, and $\phi (t)$ gives the average global phase of collective oscillations \cite{Kurths,Strogatz,kuramoto1975self}.

Several previous works have analyzed the effect that the underlying network topology has on the synchronization properties of Kuramoto oscillators. For instance, in fully connected networks, as well as in random Poissonian networks, the dynamics of the Kuramoto model yields a continuous phase transition from an incoherent state (with $R \simeq 0$) to a coherent one (with $R \simeq 1$) at the critical coupling $\sigma_c$ \cite{kuramoto1975self,strogatz2000kuramoto,acebron2005kuramoto}.  In regular lattices of dimension $d$ it has been shown that global synchronization is only possible for $d> 4$; in dimensions $2<d\leq 4$ only entrained frequency synchronization, but not phase synchronization, is observed, whereas in dimension $d\leq 2$ synchronization is not observed \cite{Choi,Chate}.  Furthermore, as mentioned above, it has been recently shown that also a regime of frustrated synchronization --akin to Griffiths phases \cite{Moretti}-- characterized by global oscillations of $R(t)$, can emerge in complex networks with hierarchical and modular structure \cite{Villegas}. This regime has been previously associated only with large-world networks, i.e. networks with a finite Hausdorff dimension \cite{Moretti}.

In what follows we illustrate that it is the network spectral dimension what mostly determines the resulting synchronization properties, so that small-world networks --with an infinite Haussdorf dimension-- of which CNMs are an example here, can also display a regime of frustrated synchronization.

\subsection{Synchronization and Frustrated Synchronization}
\begin{figure}[!h]
\center
\includegraphics[scale=0.3]{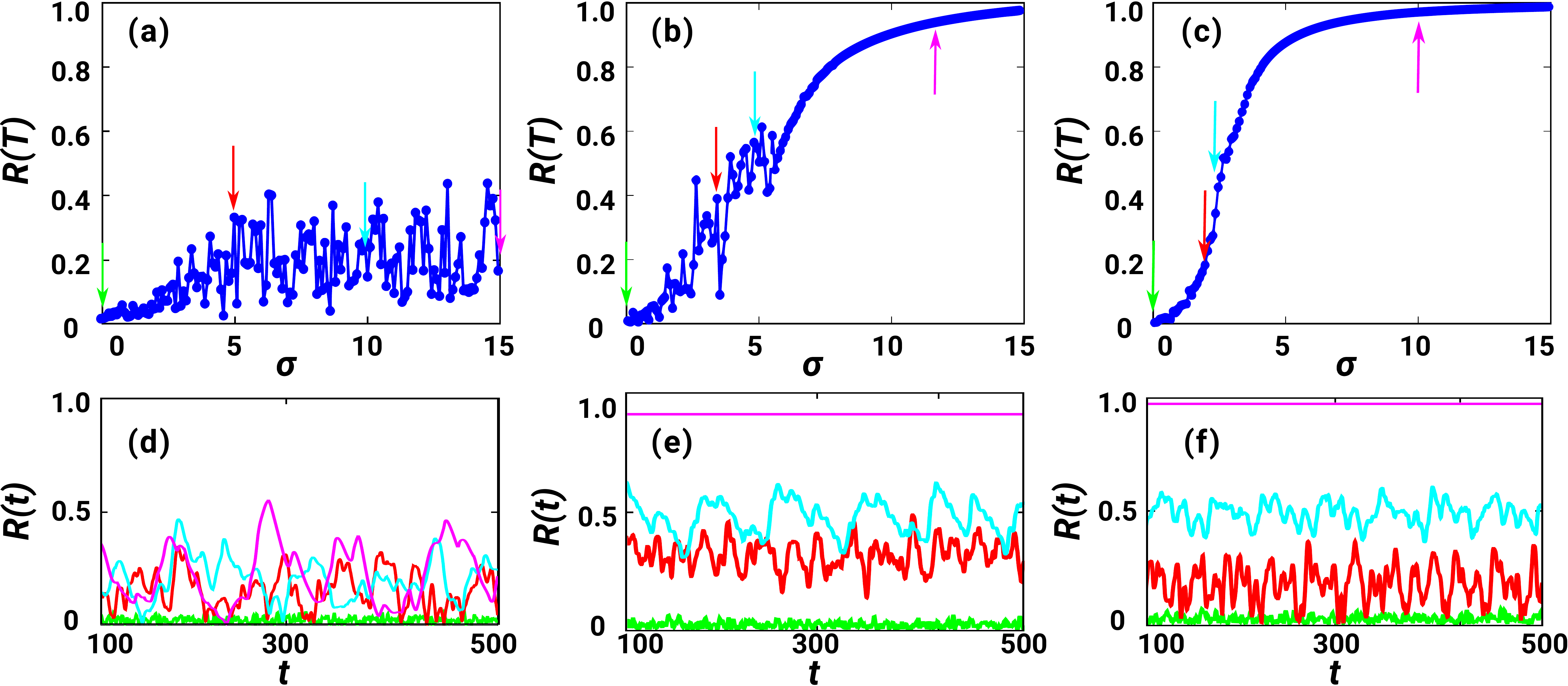} 
\protect\caption{{\bf Frustrated synchronization.}  The synchronization order parameter $R(T)$ is plotted versus the coupling strength $\sigma$ for $D=1$ (a), $D=2$ (b) and $D=3$ (c), for a single network realization of $N=1600$ nodes. Here, we have taken $T=500$ in all graphs. The arrows in the panels (a), (b) and (c) indicate the specific values of the coupling strength $\sigma$ for which we show the time series $R(t)$ in panels (d) for $D=1$, (e) for $D=2$ and (f) for $D=3$. \label{fig:sigle_run} }
\end{figure}

We have performed numerical analysis of the Kuramoto dynamics on CNMs which reveals, for a wide range of coupling values, a frustrated synchronization phase in which the global order parameter has large temporal fluctuations. In Figure $\ref{fig:sigle_run}$ we show the global order parameter $R(T)$ calculated at large times $T$ for a given fixed CNM of size $N$ and given internal frequencies $\{\omega_i\}_{i=1,2,\ldots, N}$, and for different values of the coupling $\sigma$; the same figure (bottom panels) also shows some examples of typical time series of the global order parameter $R(t)$ for $D=1,2,3$. 
Once can observe that there are regimes (of values of the coupling strength $\sigma$) in all dimensions where the synchronization dynamics does not reach a steady state with small fluctuations around a mean value.  While for $D=1$ a synchronized steady state is never reached, for $D=3$ a synchronized phase is observed (on a finite network) for large values of the coupling $\sigma$ between the oscillators. The synchronization properties in the case $D=2$ reveal an intermediate scenario with respect to the cases $D=1$ and $D=3$, with a wide region of large oscillations of the global synchronization parameter $R(T)$, i.e. frustrated synchronization.

\begin{figure}[!h]
\center
\includegraphics[scale=0.3]{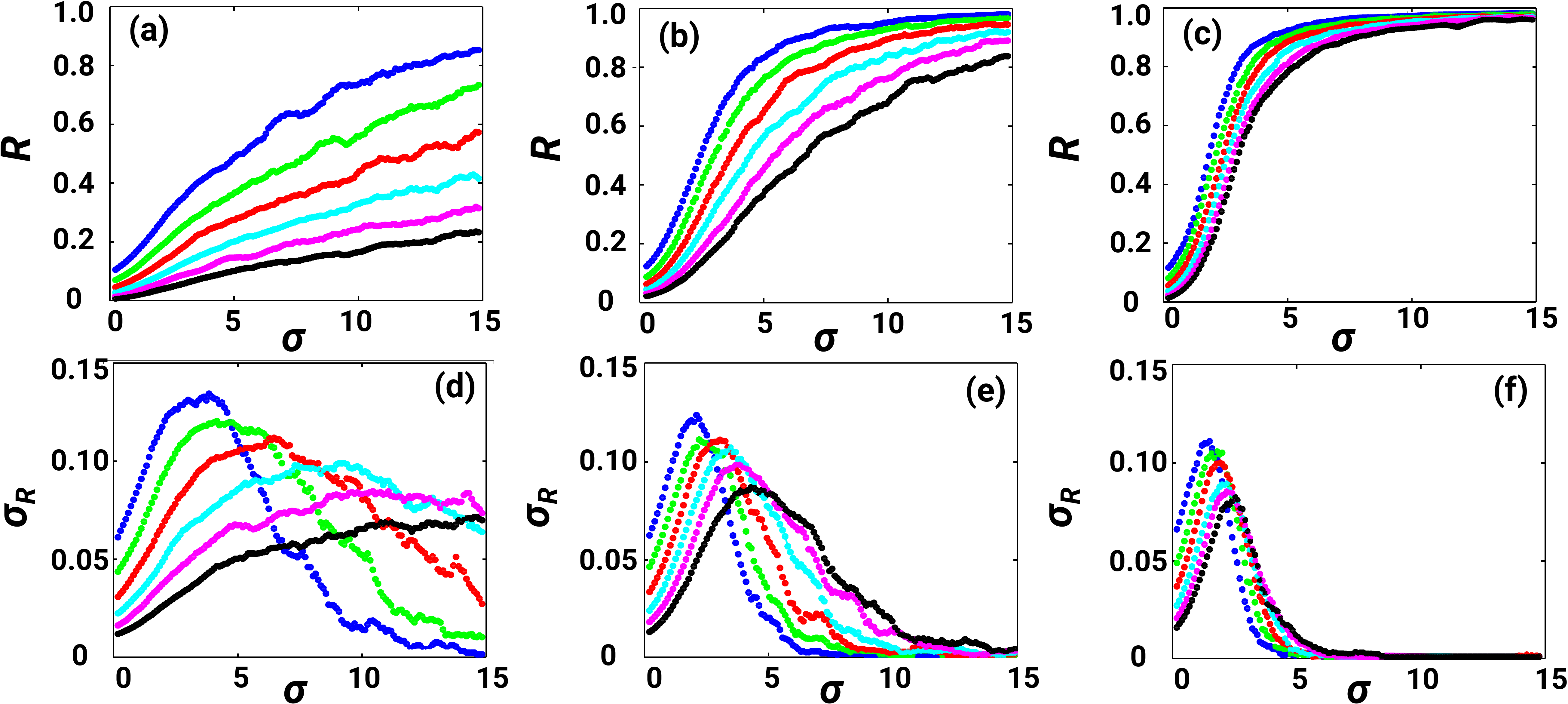} 
\protect\caption{{\bf Synchronization transition.} The synchronization order parameter $R$  and its variance $\sigma_R$ are shown  for $D=1$ (panels (a) and (d)), $D=2$ (panels (b) and (e)) and $D=3$ (panels (c) and (f)). Different lines  are drawn for  different networks sizes: $N=100$ (blue), $200$ (green), $400$ (red), $800$ (cyan), $1600$ (pink) and $3200$ (black). The data are averaged over $50$ realizations of the CNMs and the internal frequencies. A finite size scaling analysis shows decay in the synchronization order parameter with network size for every dimension (from $2$ to $4$) indicating that synchronization fades away in the infinite-network-size (thermodinamic) limit \cite{Chate}. \label{fig:high_res_2D} }
\end{figure}

In order to characterize the frustrated synchronization phase and to assess whether a true synchronization transition is observed for CNMs of $D=2$ and $D=3$, we have performed an extensive computational analysis of the considered Kuramoto dynamics averaged over different CNMs and different realizations of the internal frequencies. As a way to characterize network oscillations, in Figure $\ref{fig:high_res_2D}$ the average order parameter $R$ and its standard deviation $\sigma_R$, calculated after a transient time, are shown as a function of the coupling constant $\sigma$. The large values of the standard deviation indicate the region of the phase space in which frustrated synchronization is observed. Our finite size analysis (see Figure $\ref{fig:high_res_2D}$) reveals the strong influence of the CNM dimension on the macroscopic dynamics. 
In $D=1$ global synchronization is never achieved for large network sizes, indicating that in the large-size (thermodynamic) limit synchronization is impossible. On the other hand, synchronization in CNMs with $D=2$ and $D=3$ is possible for small networks but gets delayed to higher couplings for increasing system sizes, and a much broader regime of large fluctuations is observed in the $D=2$ case.
 
These numerical observations can be understood in connection with the spectral dimension of the corresponding CNM. In fact, by linearizing the Kuramoto dynamics, it is possible to extend  the results obtained in \cite{Choi} for regular lattices to complex networks with finite spectral dimension \cite{Daan}. These theoretical considerations reveal that for $d_S\leq 2$ networks cannot synchronize, whereas for $d_S>4$ there is always a critical value of the coupling above which synchronization is possible. Finally, for $2<d_S\leq 4$ global synchronization is not possible but an entrained synchronized state can be observed.

Interestingly, CNMs show that is possible to realize tessellations of a $D=3$ space that have spectral dimension $d_S=4$. These networks  have the critical spatial dimension for the onset of the synchronized state. This suggests that it could be possible to have marginal synchronization in three-dimensionally embedded networks.

\begin{figure}[!h]
\center
\includegraphics[scale=0.4]{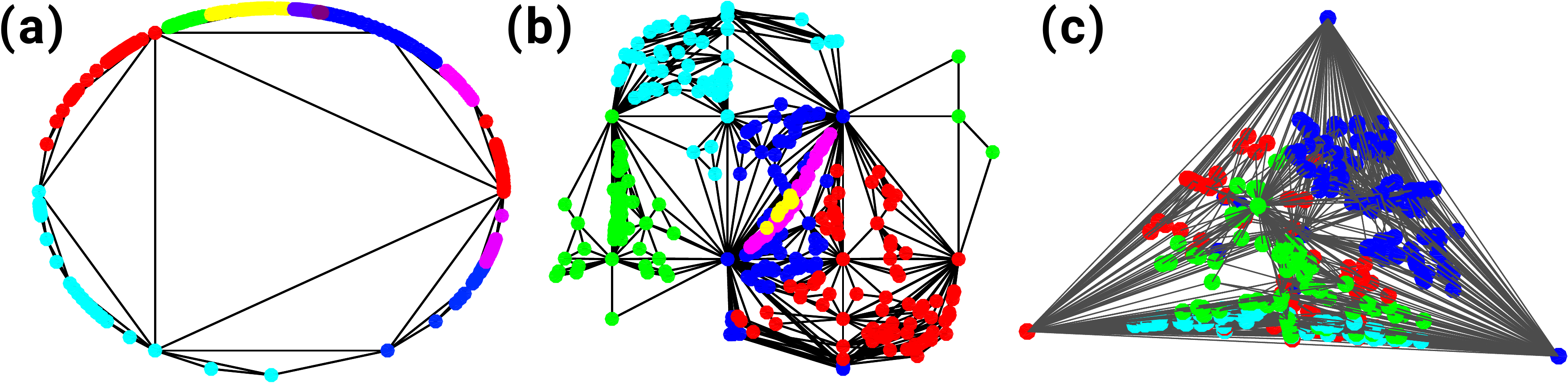}
\protect\caption{ \textbf{Geometric representation and community structure of Complex Network Manifolds.}   CNMs of $N=400$ nodes and dimension $D=1$ (a), $D= 2$ (b) and $D=3$(c) are visualized and their corresponding community structures, as detected using the Gen-Louvain algorithm \cite{Gen_louvain,Louvain}, are indicated by different colors.
$D=2$ and $D=3$ CNMs are displayed by using their holographic representation, as indicated in section {\em Network dimensions.} In $D=2$ the $2$-dimensional coordinates indicate the elevation and azimuth of the nodes on the surface of a sphere, respectively. Finally, the $D=1$ CNM is represented on a bounded $2$-dimensional manifold. The holographic $1$-dimensional representation is equivalent to this one but all links should be superimposed on the circumference. 
\label{fig:nerwork_plots} }
\end{figure}

\subsection{Spatio-temporal fluctuations of the order parameter}
Here, we investigate further the regime of frustrated synchronization by characterizing the spatio-temporal fluctuations of the order parameter.  
As a matter of fact, CNMs are characterized by significantly modular structures that can be revealed by community detection methods (see Supplementary Material). These communities define regions of the networks that include a set of nodes close in the embedding space. Additionally, these nodes are also more densely connected with each other than with other nodes of the network. In order to give a visual representation of these communities, in Figure $\ref{fig:nerwork_plots}$ we visualize single instances of CNMs in $D=1,\ D=2,\ D=3$ and plot, by employing different colors, the communities found by using a standard (Gen-Louvain) algorithm for community detection \cite{Gen_louvain,Louvain}.

\begin{figure}[!h]
\center
\includegraphics[scale=0.35]{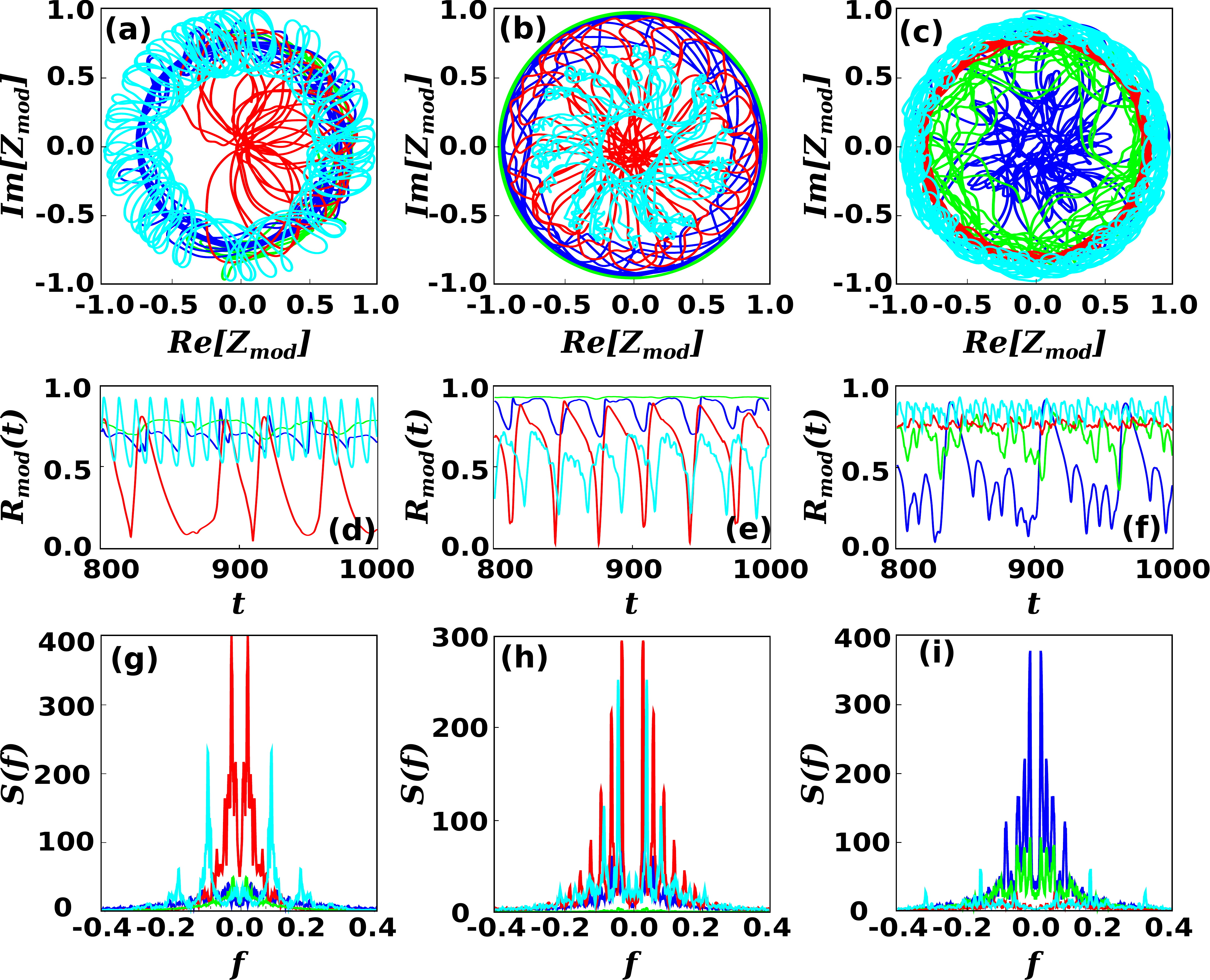}
\protect\caption{\textbf{Spatio-temporal fluctuations of the order parameter.}
Top panels show representative plots of the trajectory of the system on the phase space defined by ${\rm Re}[Z_{mod}], {\rm Im}[Z_{mod}]$ respectively for $D=1$, $\sigma=5.0$ (a), $D=2$, $\sigma=3.5$ (b) and $D=3$, $\sigma=3.0$ (c). 
The corresponding time series of $R_{mod}(t)$ is shown respectively in panels (d), (e) and (f), with the time $t$ in seconds.
Finally, the bottom panels (g), (h) and (i) show the spectral decomposition $S(f)$ of the previous time series, with the frequency $f$ in units of $10^3 [t]^{-1}$, respectively for $D=1,\ 2,\ 3$. The CNMs in this figure have size $N=800$.}
\label{fig:modular_synchro} 
\end{figure}

It is, thus, natural to explore the differences between the dynamical state of these communities. For that, we define a ``mesoscopic'' synchronization parameter as 
\bea Z_{mod,n}(t) = R_{mod}(t) e^{i\phi_{mod}(t)}=\frac{1}{|{\mathcal C}_n|} \sum_{i \in {\mathcal C}_n} e^{i \theta_i}, \eea
where ${\mathcal C}_n$ is the set of nodes in the $n^{th}$ community and $|{\mathcal C}_n|$ indicates the total number of nodes in the community. 
Here, 
\bea
R_{mod}(t)=|Z_{mod}(t)|
\eea
is a real variable taking values in the range $[0,1].$ 
Figure $\ref{fig:modular_synchro}$ displays the trajectory of $Z_{mod}(t)$ in the complex plane for some modules of exemplary cases of CNMs in $D=1,\ 2,$ and $D=3$ for coupling values that maximize $\sigma_R$ as indicated in the caption.  
In these plots, a circular trajectory describes a situation in which $R_{mod}(t)$ is constant in time, with full synchronization of the module corresponding to $R_{mod}(t)=1$.
Random trajectories around $(0,0)$ describe unsynchronized modules. Partially synchronized modules, on the other hand, may describe more complex, i.e. chaotic, trajectories.  We can distinguish between trajectories that oscillate within a circular crown of relatively large radius, in which $R_{mod}(t)$ oscillates between different states of partial synchronization, and trajectories that visit the center (unsynchronized state) too. 
Figure $\ref{fig:modular_synchro}$ also displays the time series $R_{mod}(t)$ and the spectral decomposition $S(f)$ of the temporal series $R_{mod}(t)-\Avg{R_{mod}(t)}$ for the considered exemplary modules. Interestingly, the spectral decomposition shows that, whereas some frequencies are in fact dominant, different modules can oscillate in quite different ways, indicating diverse synchronization states.

This numerical analysis reveals the relevant spatio-temporal fluctuations observed in the frustrated synchronization phase where different modules synchronize at different frequencies. Therefore, CNMs are stylized network geometries that might provide an important framework to characterize the geometrical network properties of neuronal networks and their role in brain dynamics.

\subsection{Communities and localized eigenvectors}
\begin{figure}[!h]
\center
\includegraphics[scale=0.35]{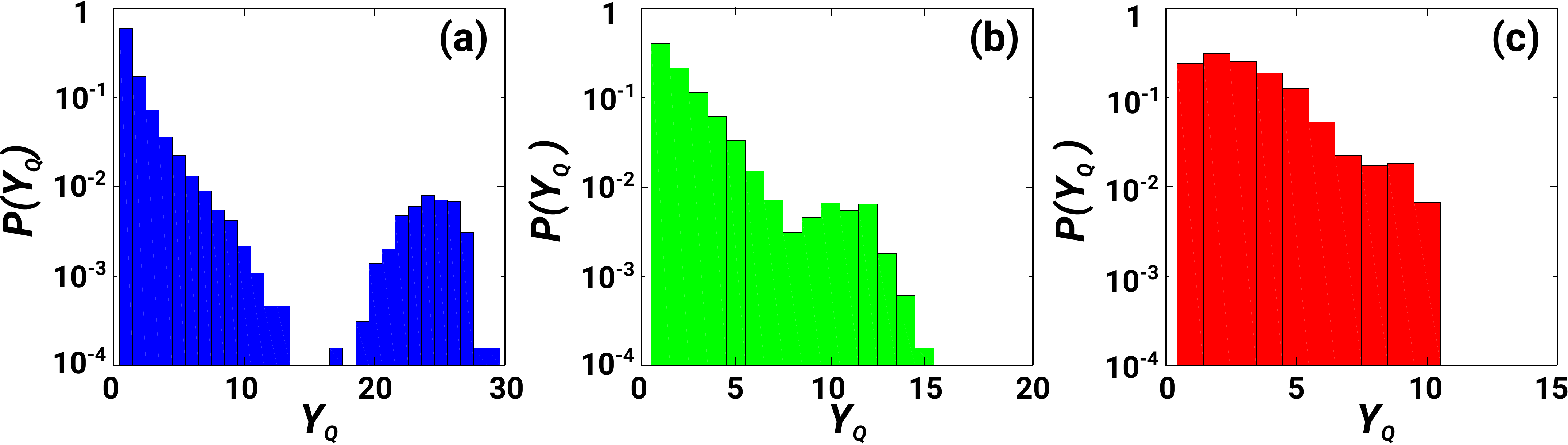}
\protect\caption{{\bf Localization of the eigenmodes.} 
The distribution $P(Y_Q)$ of the community participation ratio $Y_Q$  of a single realization of $N=6400$ nodes of a CNM is shown  for $D=1$ (panel a), $D=2$ (panel b) and $D=3$ (panel c).  The total number of communities in these networks are: 30 (panel a), 18 (panel b) and 12 panel (c). Observe that $Y$ can be larger in lower dimensions; this is because the modularity is smaller in lower dimensions and the community-detection algorithm divides the network in more communities.
\label{fig:pPR} }
\end{figure}

In order to reveal the relation between the community structure of CNMs and their spectral decomposition, we have characterized the localization properties of the eigenvector corresponding to the eigenvalue $\lambda$ on different network communities.  To this end we have evaluated for each eigenvector $\lambda$ the {\em community participation ratio} $Y_Q$, defined as 
\bea Y_{Q}=\left[\sum_{n=1}^C \left(\sum_{i\in {\mathcal C}_n} u_{i,L}^{\lambda}u_{i,R}^{\lambda}\right)^2\right]^{-1}, \eea
where ${\mathcal C}_n$ indicates the $n^{th}$ community and $C$ is the total number of communities. The community participation ratio $Y_Q\in [1,C]$ indicates the number of communities on which the eigenmode is localized. Figure $\ref{fig:pPR}$ indicates that a large number of eigenvectors are localized in one or few communities.
Thus, diverse modes are activated at different communities (and at different local coupling strengths), justifying the emergence of local patches of synchronization and overall frustrated synchronization. This therefore explains the fact that in the frustrated synchronization phase we observe a dynamics highly correlated with the modular structure of the network.

\subsection{Coarse graining of the frustrated synchronization dynamics}
\begin{figure}[!h]
\center
\includegraphics[width=0.9\columnwidth]{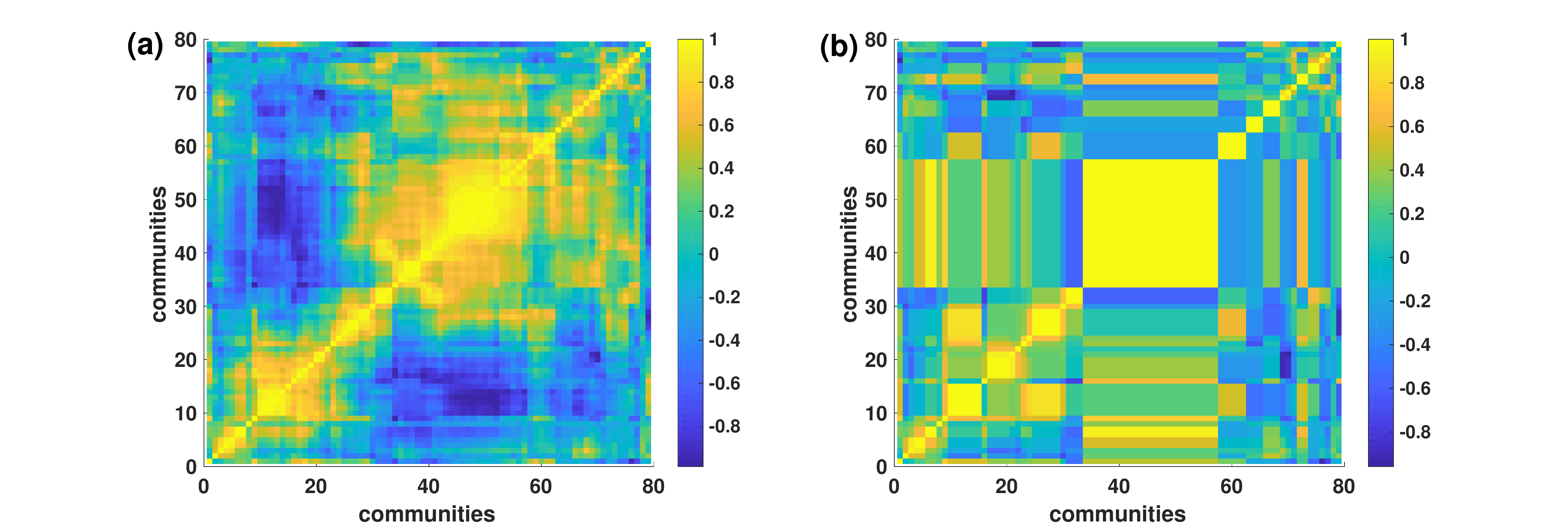}
\protect\caption{{\bf Correlations between modules and coarse graining of the frustrated synchronization dynamics.} The matrix of Pearson correlations among the dynamical state of each pair of communities is reported for a fine grained partition (panel (a)) and a coarse grained one (panel (b)) of a CNM of $N=1000$ nodes and dimension $D=2$.
The dynamical state of each community $n$ is measured using $R_{mod,n}$. The fine partition formed by 79 communities is generated by running twice the GenLouvain algorithm, yielding a modularity $Q=0.66$. The coarse grained partition has been extracted by running a single linkage clustering on the correlation matrix and cutting the resulting dendrogram in order to get 30 communities. The resulting modularity of the coarse grained partition is $Q=0.60$. \label{fig:hierarchical} }
\end{figure}

In brain research brain activity is typically measured by coarse graining the neural population dynamics at the level of macroscopic brain regions. 
The main starting point of such research is usually the extraction of a correlation matrix between the activity of different brain regions. 
Here, we investigate the role of coarse graining the frustrated synchronization dynamics on CNMs by using the local order parameter $R_{mod,n}$, which describes the internal synchronization of each module of the CNM. 
We run twice the Gen-Louvain algorithm in order to obtain a fine partition into modules of the CNM (with $D=2$), whereas retaining high modularity. We then measure the Pearson correlation between every pair of different modules (see Figure $\ref{fig:hierarchical}$(a)) to obtain a correlation matrix. Subsequently, we coarse grain further these modules (reducing the modules to about one half) by considering the aggregation generated by running a single linkage clustering to the correlation matrix (see Figure $\ref{fig:hierarchical}$(b)). We observe that the modularity of the coarse grained partition remains high, thus revealing once more the coupling between the synchronization dynamics and the network topology.
Moreover, we find that the coarse grained dynamics remains non-trivial, as revealed by a complex correlation matrix, indicating a sort of invariance under network hierarchical coarse graining.  
A similar behavior of the dynamics under coarse graining is observed in the frustrated synchronization phase of CNMs of other dimensions $D$.

\section{Conclusions and Discusion}
In conclusion, this study shows the rich interplay between network geometry and synchronization dynamics by investigating the Kuramoto model running on top of Complex Network Manifolds.
These networks define discrete manifolds in dimension $D$ with the small-world property and highly modular structure, and provide an ideal theoretical setting to explore the interplay between network geometry and brain dynamics.  
When the synchronization of CNMs of dimensions $D=3$ and $D=2$ is compared, it is observed that both network structures sustain a regime of frustrated synchronization where spatio-temporal fluctuations of the order parameter are observed. 
In this regime, network modules are characterized by different synchronization frequencies. However, finite CNMs in $D=3$ are much more favourable to sustain synchronized states than CNMs in $D=2$.  
These results help shedding light on the experimental finding that $3$D scaffolds favor neuronal network dynamics, as recorded in calcium activity experiments, with respect to $2$D geometries for neuronal cultures, and that relevant features of brain dynamics are a consequence of its $3$D topology, as experimentally observed in Ref. \cite{Torre}. 
Moreover, our study shows evidence that CNMs that can be embedded in dimension $D=3$ may have spectral dimension $d_S=4$, i.e. the critical spectral dimension for the onset of a global synchronous phase, allowing us to observe both a fully synchronized regime and a frustrated synchronization regime on a finite network. 
Also, our work reveals that non-trivial synchronization states can emerge even in small-world networks, with an infinite topological dimension.

On a wider perspective this work reveals the important role of the spectral dimension and the localization of eigenvalues in promoting the frustrated synchronization phase and opens new research lines to relate network geometry and brain dynamics.

\section*{Methods}

Computational analyses are performed in {\rm MATLAB} software. Integration of system dynamics are carried out using the {\rm ode45} function, which uses a non-stiff 4-th order integration algorithm with adaptive steps. 
Modularity analyses are performed via the Generalized Louvain algorithm \cite{Gen_louvain,Louvain}. 
The codes for generating the Complex Network Manifolds, which are equivalent to the Network Geometry with Flavor $s=-1$ \cite{NGF}, can be found in this public repository \cite{ginestrab}.
\renewcommand\theequation{{S-\arabic{equation}}}
\renewcommand\thetable{{S-\Roman{table}}}
\renewcommand\thefigure{{\arabic{figure}S}}
\setcounter{equation}{0}
\setcounter{figure}{0}
\setcounter{section}{0}

\section*{Acknowledgements}
We are grateful for financial support from Spanish MINECO (under grant FIS2013-43201-P; FEDER funds) and from \textsf{"Obra Social La Caixa"}. A.P.M. acknowledges the kind hospitality of the School of Mathematical Sciences at Queen Mary University of London where this work started. The authors acknowledge interesting and fruitful discussions with M.A. Mu\~noz. 

\section*{Author contributions statement}
A.P.M.,   J. J. T. and G.B. designed the research, conducted the research and wrote the manuscript, A.P.M. performed the simulations and prepared the figures.

\section*{Additional information}
Codes are available upon request to the authors.
The authors declare no competing financial interests.

\clearpage

\begin{center}
\textbf{\LARGE{Supplementary Material for ``Complex Network Geometry and Frustrated Synchronization''}}

\vspace{5mm}

In this Supplementary Material we provide additional information about the structure of Complex Network Manifolds regarding its degree distribution and its small-word character. 

\end{center}

\vspace{5mm}

\section{Degree distribution and Hausdorff dimension of Complex Network Manifolds}

The degree distribution $p(k)$ of Complex Network Manifold \cite{CQNM,NGF} is exponential for dimension $d=2$ and scale-free for $d>2$. The exact asymptotic expression has been derived in Ref. \cite{Hyperbolic} and is given for $d=2$ by
\begin{equation} \label{eq2d}
p(k)=\frac{1}{d+1} \left(\frac{2}{3}\right)^{k-d},
\end{equation}
with $k\geq 2$ whereas for  $d>2$ it is given by  
\begin{equation} \label{eq34d}
p(k)=\frac{d-1}{2d-1} 
\frac{\Gamma \left[ (1+(2d-1)/(d-2) \right]}{\Gamma \left[ d/(d-2) \right]}
\frac{\Gamma \left[k-d+d/(d-2) \right]}{\Gamma \left[k-d+1+(2d-1)/(d-2) \right] }, 
\end{equation}
with $k\geq d$. Therefore, for $d>2$ Complex Network Manifolds  are scale-free with a power-law scaling 
\begin{equation}
p(k) \approx k^{-\gamma},
\end{equation}
valid for $k\gg 1$, and power-law exponent $\gamma$ 
\begin{equation}
\gamma= 2 + \frac{1}{d-2}.
\end{equation}
Therefore, for $d=3$ we obtain  $\gamma = 3$ and  for $d=4$ we obtain $\gamma = 5/2=2.5$. 
In Fig. 1S (a) we show the agreement between the analytic expression (dashed lines) and the computational results (data points).

 The  logarithmic scaling of the average shortest (hopping) distance $\ell$ between the nodes of the network with the network size $N$, is known to reveal the small-world \cite{SW} nature of a network.
 By investigating numerically the scalling of $\ell$ with $N$ we show that  Complex Network Manifolds are small world.
 Our result are reported in   Fig.1S (b) where  points represent data from numerical simulations of Complex Network Manifold of dimension $d$, whereas the solid lines stand for the best logarithmic fit, as given by 
\bea
\ell = a_d \log(N) + b_d.
\label{eq:log}
\eea

The parameters from the fit are shown in the Table $\ref{table_SW}$,  and they clearly indicate that the Complex Network Manifolds of higher dimension $d$ have a average shortest distance that grows always logarithmically with the network size $N$ but with different constant prefactor $a_d$.

\begin{table}[htb!]
\begin{center}
\begin{tabular}{|c|c|c|c|}
\hline 
$d$ & $a_d$ & $b_d$ & $R^2$\tabularnewline
\hline 
\hline 
$2$ & $2.93(3)$ & $-1.45(9)$ & $0.983$\tabularnewline
\hline 
$3$ & $1.32(2)$ & $0.17(4)$ & $0.964$\tabularnewline
\hline 
$4$ & $0.78(1)$ & $0.79(4)$ & $0.954$\tabularnewline
\hline 
\end{tabular}
\par\end{center}
\caption{Fitted parameters $a_d$ and $b_d$ determining the logarithmic growth of the average shortest (hopping) distance $\ell$ of Complex Network Manifolds in dimension $d$ according to Eq. $(\ref{eq:log})$.}
\label{table_SW}
\end{table}

\begin{figure}[htb!]
\center
\includegraphics[scale=0.5]{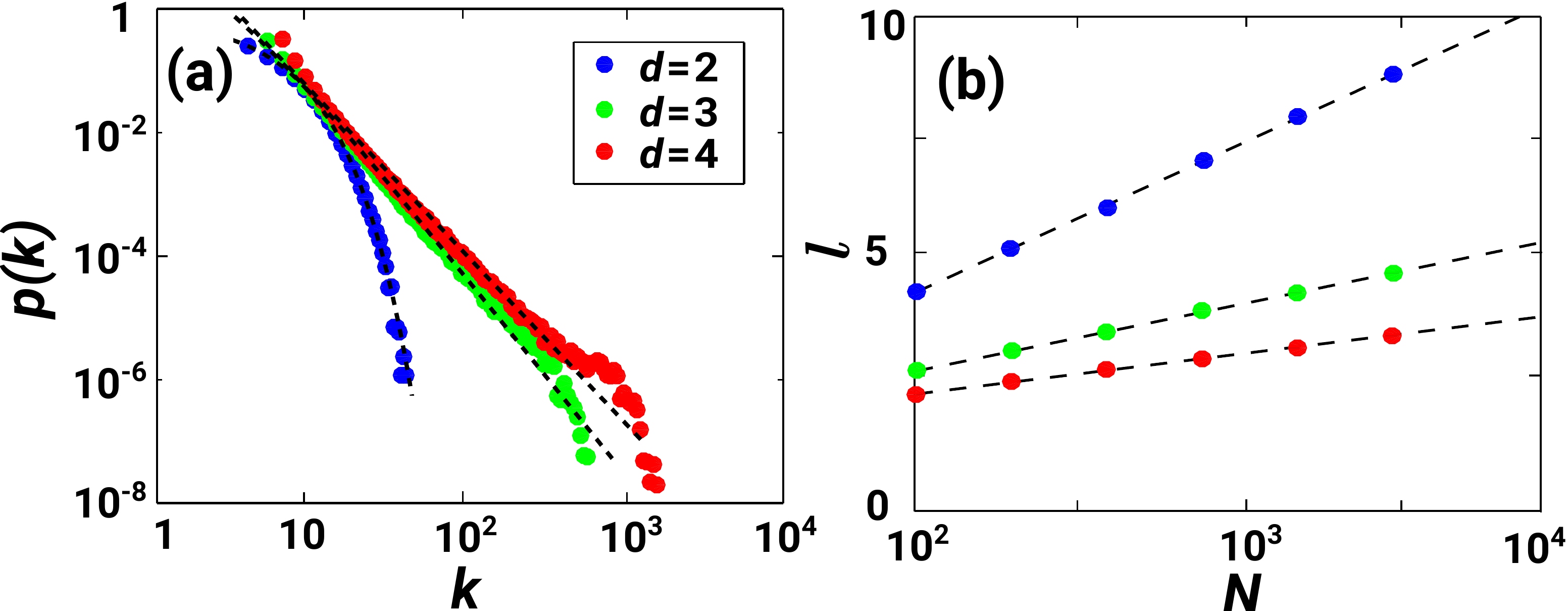} 
 \label{fig:NGF_netws} 
 \protect\caption{\textbf{Degree distribution and small-word properties of Complex Network Manifolds.} The degree distribution $p(k)$ of the Complex Network Manifolds  of $N=6400$ nodes and dimensions $d=2, 3$ and $4$ is plotted in panel (a). Points represent results from numerical simulations whereas dashed lines stand for the analytical result as given by eq. \ref{eq2d} and \ref{eq34d}.  The network diameter $D$ is plotted versus the network size $N$ for dimensions $d=2,\ 3,\ 4$ in panel (b). Data points are from simulation results whereas dashed lines correspond to the  logarithmic fit. Numeric results have been averaged over 100 network realizations in both plots.}
\end{figure}


\begin{thebibliography}{99}
\bibitem{Note}
  Previous studies on frustrated synchronization on complex networks have shown that whereas the particular choice of the distribution can affect quantitatively the phase transition between an asynchronous and a synchronous phases, its qualitative properties are quite robust \cite{Villegas}.
\end{thebibliography}
\end{document}